\documentclass[aps,prl,twocolumn,nopacs,superscriptaddress]{revtex4}
\usepackage{graphicx}
\usepackage{verbatim}
\usepackage{bibunits}
\begin{document}
\begin{bibunit}

\title{Electron behavior in topological insulator based P-N overlayer interfaces}

\begin{abstract}

Topological insulators (TIs) are novel materials that manifest spin-polarized Dirac states on their surfaces or at interfaces made with conventional matter. We have measured the electron kinetics of bulk doped TI Bi$_2$Se$_3$ with angle resolved photoemission spectroscopy while depositing cathodic and anodic adatoms on the TI surfaces to add charge carriers of the opposite sign from bulk dopants. These P-N overlayer interfaces create Dirac point transport regimes and larger interface potentials than previous N-N type surface deposition studies, revealing unconventional Rashba-like and surface-bulk electron interactions, and an unusual characteristic distribution of spectral weight near the Dirac point in TI Dirac point interfaces. The electronic structures of P-N doped topological interfaces observed in these experiments are an important step towards the understanding of solid interfaces with topological materials.

\end{abstract}

\author{L. A. Wray}
\affiliation{Department of Physics, Joseph Henry Laboratories, Princeton University, Princeton, NJ 08544, USA}
\affiliation{Advanced Light Source, Lawrence Berkeley National Laboratory, Berkeley, California 94305, USA}
\author{M. Neupane}
\author{S.-Y. Xu}
\author{Y.-Q. Xia}
\affiliation{Department of Physics, Joseph Henry Laboratories, Princeton University, Princeton, NJ 08544, USA}
\author{A. V. Fedorov}
\affiliation{Advanced Light Source, Lawrence Berkeley National Laboratory, Berkeley, California 94305, USA}
\author{H. Lin}
\author{S. Basak}
\author{A. Bansil}
\affiliation{Department of Physics, Northeastern University, Boston, MA 02115, USA}
\author{Y. S. Hor}
\affiliation{Department of Physics, Missouri University of Science and Technology, Rolla, MO 65409}
\affiliation{Department of Chemistry, Princeton University, Princeton, NJ 08544, USA}
\author{R. J. Cava}
\affiliation{Department of Chemistry, Princeton University, Princeton, NJ 08544, USA}
\author{M. Z. Hasan}
\affiliation{Department of Physics, Joseph Henry Laboratories, Princeton University, Princeton, NJ 08544, USA}

%\pacs{PACS numbers~:~~74.20.Mn, 74.72.-h, 71.45.Lr, 74.50.+r }

%\pacs{}

\date{\today}

\maketitle

\begin{figure}[t]
\includegraphics[width = 8.5cm]{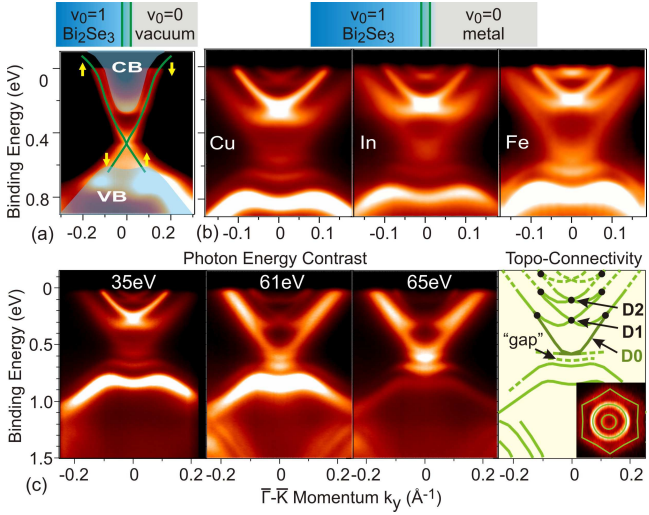}
\caption{{\textbf{Topological connectivity in a TI-metal P-N interface}}: (a) The (blue) bulk and (green) surface band structure of Bi$_2$Se$_3$ is traced above ARPES measurements of band structure with N-type bulk doping. A diagram above the ARPES image shows topological quantum numbers and the location of the topological surface states. (b) ARPES images are shown for cathode (Cu-, In- and Fe-) deposited P-type Bi$_2$Se$_3$, measured after saturating the surface so that the binding energy of surface states no longer changed with further deposition. The Z$_2$ topological quantum number $\nu_0$ of interfaced materials is indicated above. (c) The surface states of Cu-deposited Bi$_2$Se$_3$ resonate at different photon energies. Measurements have a $\lesssim$1nm penetration depth into the crystal \cite{univCurve}. An integrated comparison reveals the band structure traced at right, with points at which bands converge labeled in black, including new Dirac points labeled D1 and D2. The Fermi surface is shown as an inset. States that evolve adiabatically from the original upper Dirac cone (D0) are drawn in a darker hue.}
\end{figure}

Bismuth based topological insulators (TIs) are the first experimentally realized three dimensional topologically ordered bulk solids, and feature robust, massless spin-helical conducting surface states that appear at any interface between a TI and normal matter that lacks the topological order \cite{Intro,TIbasic,TopoFieldTheory,DavidNat1,DavidScience,Chen,MatthewNatPhys}. This topologically defined surface environment has been theoretically identified as a promising platform for observing a wide range of new physical phenomena, and possesses ideal properties for advanced electronics such as spin-polarized conductivity and suppressed scattering \cite{DavidScience,MacDonaldKerr,FuNew,FerroSplitting,FuHexagonal,ExcitCapacitor,ZhangDyon,KaneDevice,DavidNat1,MatthewNatPhys,ZhangPred,Chen,YazdaniBack,WrayCuBiSe,WrayFe}. Developments from the last year have demonstrated that it is possible to fabricate nanodevices using Bi$_2$Se$_3$ and pass current through the surface states \cite{topGate,sacebeDevice}, but have also highlighted the fact that topological insulators are a new state of matter distinct from normal metals and insulators, and their properties in a junction/interface setting are not known or understood \cite{WrayCuBiSe,WrayFe,HofmannRashba2DEG,PanNN,DamascelliRashba,BeniaH2O}.

In this Letter, we explore this critically important question by using cathode surface overlayers with weak and strong spin-orbit interactions (Cu and In) and anodic nitrogen dioxide in tandem with control of bulk doping in TI Bi$_2$Se$_3$
from P-type to N-type charge carrier regimes for the first time to generate a range of topological insulator P-N interface scenarios that are relevant for device development \cite{KaneDevice,topGate,sacebeDevice}.
The P-N scenario allows us to access the novel aspects of the physics not observed in previous studies limited to N-N scenarios. Our experimental results reveal diode-like configurations that can manifest a gap in the interface electron density near a topological Dirac point and systematically modify the interface Dirac velocity, allowing far reaching control of helical electrons at the interface. In addition, we show that our data are consistent with first principles numerical models suggesting that the Rashba-like spin-orbit effect in biased TI junctions derives directly from the symmetry inversion of topological order, and is instrumental in maintaining the topological connectivity of the surface 2D band structure at large interface potentials.

\begin{figure}[t]
\includegraphics[width = 8cm]{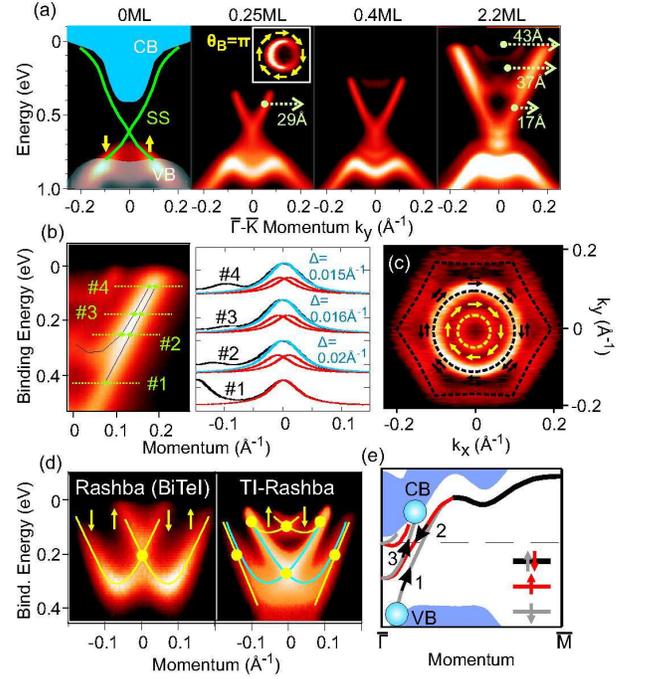}
\caption{\label{fig:PNfig}{\bf{Topological basis for an unconventional Rashba effect}}: (a) Surface deposition of copper is shown on bulk P-type Bi$_2$Se$_3$ (0.5$\%$ Ca doped as Bi$_{1.995}$Ca$_{0.005}$Se$_3$), with approximate pre-deposition electronic structure traced on the first panel and energies shifted to align surface band features between panels. The mean free path estimated from inverse momentum width is labeled in angstroms for surface conduction bands. (inset) The Fermi surface is shown for 0.25ML of Cu deposition with spin orientations labeled by arrows. (b) (black) Momentum dispersion curves of the upper D0 Dirac cone and D1 surface state are shown with (blue) fit curves that are composed of (red) $\Gamma$=0.06$\AA^{-1}$ Lorentzians separated by `$\Delta$'. (c) The Fermi surface after 2.2ML Cu deposition is shown with numerically predicted spin orientations. (d) (left) Large Rashba splitting in the surface states of topologically trivial BiTeI is compared with (right) the strikingly different Rashba-like effect in topologically ordered surface electrons. Locations at which bands intersect or qualitatively appear to converge are labeled with yellow circles. (e) The topological connection between the valence band (VB) and conduction band (CB) is traced with black arrows on a DFT-based simulation using a coarse-grained band bending potential from Ref. \cite{WrayFe}.}
\end{figure}

The surface state of TI Bi$_2$Se$_3$ crystals cleaved in ultra high vacuum (UHV) is a singly degenerate Dirac cone that spans a bulk semiconductor band gap of $\Delta$$\gtrsim$250meV (Fig. 1(a)), with a Fermi level that can be moved through the gap by bulk dopants such as Ca.
% \cite{ChenMatthewNatPhys,ZhangPred,WrayCuBiSe}.
Evaporating strong electron donors on P-type bismuth selenide moves the surface chemical potential across the band gap, achieving a conventional diode like interface voltage of up to V$_d$$\sim$0.7V in our data from a sub-nanometer surface layer that is permeable to photoexcited electrons \cite{univCurve}. Although metal overlayers are used in this study to accommodate the shallow penetration depth of photoemission, a thicker and less metallic semiconductor overlayer is likely be more practical for conventional P-N device applications, in which case a gating potential might be required to achieve (or exceed) the interface potentials considered here. The cleaved Se$^{2-}$ surface of Bi$_2$Se$_3$ is maintained at low temperature (generally T$<$15K), which is critical to provide a stable and chemically neutral substrate for these adatoms, and angle resolved photoemission spectroscopy (ARPES) performed at Advanced Light Source beamlines 12 and 10 is used to map the electronic states with energy resolution better than 20 meV. Measurements with scanning tunneling microscopy and low energy electron microscopy indicate that atoms such as Cu and Fe do not clump on the surface, but rather grow in layers \cite{HorSC,LEEM}.  Deposition depth is presented in units of monolayers (ML) to approximately represent the number of deposited atoms per hexagonal unit cell of the surface, based on rough linear assumptions outlined in the online Supplemental Material \cite{SMo}.

After an estimated 1-3ML of cathode deposition on hole doped (P-type) Bi$_2$Se$_3$ the Fermi level ceases to move as more adatoms are deposited. The topological surface state is dramatically changed by the nanoscale P-N interface, manifesting new intersecting Dirac cones and a gap in the surface density of states near the original (D0) Dirac point after the surface state binding energy has been moved by 0.5-0.7eV (Fig. 1(b)), corresponding approximately to the chemical interface potential in numerical simulations (V$_d$). Contrasting photoemission images measured with different incident energies reveals the connectivity of energy/momentum band contours traced in Fig. 1(c), which must be known to understand topological order at a TI-metal junction. Related studies of N-N overlayer nanoscale interfaces are limited to smaller interface biases (V$_d$$\lesssim$0.4V) \cite{HofmannRashba2DEG,PanNN,DamascelliRashba,BeniaH2O}, and chemical potentials far from the Dirac point (D0) where topological phenomena are optimized. In this present work, our main focus is the non-magnetic P-N overlayer interfaces studied for the first time, from which we observe the topological connectivity of interface states which may be seen more clearly than in earlier studies performed with smaller interface voltages.

The incremental evolution leading to the appearance of multiple new Dirac cones \cite{diracExplain} on cathode-deposited P-type Bi$_2$Se$_3$ is shown in Fig. \ref{fig:PNfig}(a). The build-up of positively charged adatoms first donates electrons to the original Dirac cone of freshly cleaved Bi$_2$Se$_3$, and then gradually binds four new extrinsic surface states that intersect in two new Dirac points (D1 and D2) at the Brillouin zone (BZ) center. Tracing these new states from their points of intersection in the BZ center, we observe that they exchange partners and appear to merge with a different surface band at momenta larger than k$\sim$0.1$\AA^{-1}$. This type of connectivity is also proposed in our previous work using Fe-deposition \cite{WrayFe}, but experimental evidence in the present study is stronger due to the Cu-deposited interface achieving sharper bands that are visible over a larger energy range, and due to the absence of potential magnetic interactions. Momentum distribution curves of the combined D0/D1 feature show that it becomes narrower approaching the Fermi level. Fitting the two bands with $\Gamma$=0.06$\AA^{-1}$ Lorentzians suggests that they are $\sim$0.02$\AA^{-1}$ apart at 250meV binding energy and $\sim$0.015$\AA^{-1}$ apart at 75meV binding energy (Fig. \ref{fig:PNfig}(b)). These fits only constitute an estimation, as the exact broadening function required to fit these bands is not well known due to the parabolicity of the D1 band and other factors discussed below. The uncertainty in band separation due to the choice of broadening function may be quite large for cut $\#$2 \cite{SMo}. The combined feature has a velocity of v$_D$=2.5$\pm$0.1eV-$\AA$, and the velocities of the two bands appear to differ by just $\lesssim$7$\%$ over the 0.2eV range approaching the Fermi level, in spite of the fact that one of the bands is induced by a surface potential and the other is the seemingly unrelated topological insulator upper Dirac cone.

The resulting Fermi surface after heavy deposition (Fig. \ref{fig:PNfig}(c)) is made up of three rings, of which the outer two are doubly degenerate within intrinsic width resolution. Doubly degenerate (or near-degenerate) Fermi surface rings cannot carry a non-zero Berry's phase \cite{DavidScience}, meaning that the Berry's phase of $\Theta_B$=$\pi$ by which a topological insulator interface is identified is obtained from the innermost singly degenerate Fermi surface ring. The outermost ring is found at k$\sim$1.7-2.0$\AA^{-1}$, and is hexagonally warped due to the crystal symmetry \cite{FuHexagonal}. We refer to all of these surface bands as topological surface (or interface) states, because they manifest at the surface of the material and all five Dirac bands must be observed to measure the $\Theta_B$=$\pi$ Berry's phase of a topological interface. This terminology does not preclude the accuracy of other terms (e.g. 2D electron gas, 2DEG), and the role of extrinsic surface states in topological surface band conductivity must be evaluated on a case-by-case basis for different interfaces as discussed below.

As compared to the conventional Rashba states which can be adiabatically removed from the bulk band-gap, the TI Rashba states observed here are inherently required by the nontrivial topology of the TI system. As shown in Fig. 2(e), in first principles simulations, a substantial P-N junction voltage will make the original surface states no longer connect the bulk valence band to the bulk conduction band. Therefore, within band-bending models the TI Rashba-like states have to appear in order to make a zig-zag linkage of states that connect across the bulk bandgap, which satisfies the topology. Unlike the true Rashba effect illustrated in Fig. 2(d,left), the spin splitting found in our models of Bi$_2$Se$_3$ is caused by the bulk symmetries that make the crystal a TI, which occur in a small symmetry inversion region at k$\lesssim$0.1$\AA^{-1}$ in the BZ center. We note that the conventional Rashba effect causes weak splitting of the bands at all momenta and forbids exact double degeneracy away from the Kramers high symmetry points, but is too weak to be observed by eye within the plot. Therefore, interface states are nearly-doubly-degenerate throughout most of the BZ, as seen from the thick black line at k$\gg$0.1$\AA^{-1}$ in Fig. 2(e).

\begin{figure}[t]
\includegraphics[width = 8cm]{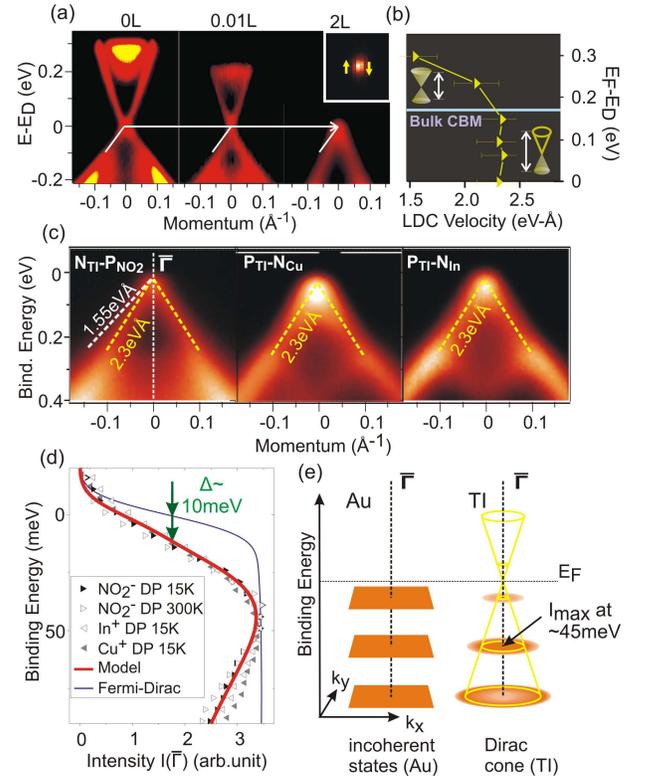}
\caption{\label{fig:InFig}{\bf{Universal Dirac point kinetics and lineshape}}: (a) Deposition of anode NO$_2$ on N-type as-grown Bi$_2$Se$_3$, with the original slope of the lower Dirac cone overlaid on each panel. An inset shows the Dirac node Fermi surface after 2L deposition. (b) The lower Dirac cone band velocity is plotted as a function of chemical potential above the Dirac point as NO$_2$ is deposited. (c) Photoemission images of bulk P-type and N-type samples with chemical potential set at the Dirac point via surface deposition of negatively and positively charged molecules, respectively (NO$_2$, Cu and In). The lower Dirac cone velocity of pristine pre-deposition as-grown Bi$_2$Se$_3$ is traced in white. (d) ARPES intensity as a function of binding energy is plotted at the $\overline{\Gamma}$-point for Dirac point interfaces shown above. A constant density of states curve cut off by the Fermi function is shown in blue, resembling polycrystalline gold, and a red curve shows simulated intensity obtained by broadening an ideal v$_D$=2.3eV-$\AA$ Dirac cone with Lorentzian parameters $\Gamma_k$=0.0275$\AA^{-1}$ and $\Gamma_E$=0.012eV. (e) Small panels illustrate the occupied electronic states of Bi$_2$Se$_3$ and polycrystalline Au as a function of binding energy, demonstrating that momentum broadening can create a peak in intensity at finite binding energy.}
\end{figure}

Topological insulators are among the very few material systems with band structures that can be manipulated to achieve transport defined by a Dirac point, and most proposed applications for TIs derive from the behavior of low density spin-helical Dirac surface electrons. In Fig. \ref{fig:InFig}(c) we present three realizations of P-N overlayer nanoscale interfaces in which the Bi$_2$Se$_3$ surface state Dirac point is brought to the Fermi level where it will shape transport properties. The surface chemical potential of an N-type sample (as-grown Bi$_2$Se$_3$ with [Se]$^{2+}$ defects) is lowered by depositing NO$_2$$^-$, and a P-type sample (Bi$_{1.995}$Ca$_{0.005}$Se$_3$) is modified through deposition of Cu$^+$ and In$^+$. Our data show that lowering the surface chemical potential of an N-type sample to the Dirac point has the effect of dramatically increasing the Dirac velocity from v$_D$=1.55$\pm$0.2eV$\cdot\AA$ to 2.30$\pm$0.2eV$\cdot\AA$, observed from the lower Dirac cone band slope traced in Fig. \ref{fig:InFig}(a). The velocity of lower Dirac cone electrons changes rapidly after deposition begins, but stabilizes after the surface chemical potential falls below the bulk conduction band minimum 0.17$\pm$0.02eV above the Dirac point (Fig. \ref{fig:InFig}(b)). The Dirac point energy is estimated from the point at which the bands are narrowest along the momentum axis, and the large error bar of $\pm$0.2eV$\cdot\AA$ is assigned from the deviation between slope estimates over 100meV and 300meV windows beneath the Dirac point. With the Fermi level inside the band gap, bulk conduction electrons are strongly repelled from the surface and are not observed within 100meV of the Fermi level in the ARPES image.

When the chemical potential lies inside the bulk electronic gap as measured by surface-sensitive photoemission, the lower Dirac cone velocity has the same value of 2.3$\pm$0.2eV$\cdot\AA$ for both N-type and P-type samples, which is strikingly inconsistent with expectations based on the conventional Rashba effect. The Hamiltonian for a conventional Rashba effect induced by an external potential is given as $H_R = \alpha_R(\vec{k}\times\hat{z})\cdot\vec{\sigma}$, where $\vec{\sigma}$ represents the Pauli matrices and $\alpha_R\propto E_Z$ is the Rashba parameter which is proportional to the z-axis electric field that breaks parity symmetry. This Hamiltonian has a nearly identical form to the linear term defining the Bi$_2$Se$_3$ Dirac velocity, $H_{TI} = v_D(\vec{k}\times\hat{z})\cdot\vec{\sigma}$ \cite{FuHexagonal}, where $v_D$ is the Dirac velocity. The lowest order combined effect of conventional Rashba interactions and topological surface physics is therefore to change the Dirac velocity to a new value of $v_D'=\left|v_D+c*E_Z\right|$, where `c' is a constant. From this, one would expect the Dirac velocity change in opposite directions when the surface state is exposed to adatoms with opposite charge. The fact that the surface velocity does not show this sort of dependence is consistent with our numerical result in Fig. 2(e) that the conventional Rashba effect is actually very weak in surface deposited Bi$_2$Se$_3$, and suggests that other factors such as bulk carrier density near the surface may be significant in creating the band instability.

Because the lower Dirac cone velocities are roughly identical for all of these Dirac point interfaces, it is convenient to go further in comparing their spectral line shapes. As shown in Fig. \ref{fig:InFig}(d), the energy dispersion curve (EDC) in the Brillouin zone center is similar for all of the photoemission measurements with the Dirac point at the Fermi level. These curves show a soft tail extending to large binding energy, a broad peak at 40-50meV binding energy, and a much shallower slope approaching the Fermi level than would be expected from the Fermi Function convoluted with Gaussian resolution broadening (blue curve). There is no simple way to obtain such an energy contour with only Fermi function and energy-axis broadening effects on the Dirac node. Because the energies of bulk electronic states are biased significantly in opposite directions for P$_{TI}$-N (e.g. Cu or In deposition) and N$_{TI}$-P (NO$_2$ deposition) interfaces, it is unlikely that resonance with bulk electronic states is responsible for any of these features that are shared between all of the curves.

To understand where this lineshape may come from, we briefly consider the asymmetrical roles of momentum-space and energy-axis intrinsic widths in defining the ARPES spectrum from 2D Dirac bands. Applying momentum broadening to an ideal 2D Dirac cone reshapes spectral intensity in the BZ center to I(E)$\propto$ $\frac{\left|E\right|}{(E\times v_D)^2-(\Gamma_k/2)^2}$ where `E' is energy relative to the Dirac point and $\Gamma_k$ is the Lorentzian momentum broadening term. We have chosen momentum and energy broadening parameters of $\Gamma_k$=0.0275$\AA^{-1}$ and $\Gamma_E$=0.012eV such that the resultant broadening vector is approximately normal to the Dirac cone in the 2+1 dimensional momentum/energy space and corresponds with the intrinsic width prior to broadening from experimental resolution. The characteristic EDC lineshape that emerges from momentum broadening is not particularly sensitive to the detailed values of $\Gamma_k$ and $\Gamma_E$, so long as $\Gamma_k$ is taken to be the dominant component determining linewidths. The resulting EDC spectral function has a peak $\sim$40meV away from the Dirac node (Fig. \ref{fig:InFig}(e)), and closely matches the experimental result when the T=15K Fermi function and weak energy-axis broadening are also applied (red curve in Fig. \ref{fig:InFig}(d)).

The observation of a line shape representing strong momentum-space broadening is consistent with recent studies which suggest that microscopic surface domains on topological insulators have offset momentum symmetries \cite{YazdaniKbrd}, causing them to be broadened in momentum space rather than along the energy axis when one performs a spatially averaging measurement such as the photoemission measurements discussed here (beam spot area A$>$200$\times$200$\mu m^2$). Broadening in momentum space may also be understood as following from the same justification as broadening in energy, because electrons have finite spatial coherence lengths, and momentum (e.g. `k$_x$') and spatial variables (e.g. `x') are conjugate to one another. Typical bands approaching the Fermi level resemble pseudo-Voigt functions, cut by the step-like Fermi Dirac function, with intrinsic widths in energy and momentum ($\Gamma_k$,$\Gamma_E$) that are related by the group velocity $v_g$ as $\Gamma_k$=$\Gamma_E$/$v_g$. In most cases, there is no easy way to distinguish between broadening that is `applied' along the energy or momentum axis, as the resulting distributions of spectral intensity are nearly identical.

However, close to a Dirac point singularity, applying a broadening convolution along the energy axis causes very different lineshapes than broadening bands along the momentum axis, and the relation $\Gamma_k$=$\Gamma_E$/$v_g$ no longer has a rigorous physical justification. States with significant nanometer-scale spatial `texture' such as those observed by STM in Ref. \cite{YazdaniKbrd} provide one physical motivation for asymmetrical broadening, as the lifetime of these states can in principle be quite long (small $\Gamma_E$) even when their ensemble distribution in momentum space is very broad (large $\Gamma_k$). It is well known that the momentum-integrated electronic density of states goes to zero at a Dirac point in the absence of self energy broadening. Momentum broadening ($\Gamma_k$) does not effect the distribution of electronic states along the energy axis, and thus allows the density of states at the Dirac point to remain zero. Energy-axis broadening ($\Gamma_E$, imaginary self energy) causes the momentum-integrated density of states at the Dirac point energy to become non-zero, and to approach the Dirac point energy in a parabolic rather than linear fashion.

Interfaces between conventional semiconductors and metals can be understood through the well known phenomenon of band bending, in which the conduction and valence bands of each material bend to line up at the interface. In the case of a TI interfaced with a normal metal, the inversion of symmetries between the topological insulator valence and conduction bands makes it impossible for the bulk band structures to connect in such a simple way \cite{TIbasic}. To understand the unusual interface properties observed in our data, we have created a Green's function implementation of the experimentally-based \textbf{k}$\cdot$\textbf{p} model in Ref. \cite{FuNew,FuHexagonal} to numerically simulate junctions at the surface of a \emph{semi-infinite topologically ordered} Bi$_2$Se$_3$ slab based on experimentally measured bulk electron kinetics from Ref. \cite{WrayCuBiSe,fisherBending}. Using a screening potential that matches the observed change in topological surface state electron energies in the Bi$_2$Se$_3$-cathode P-N interface ($\Delta_\mu$=0.7eV), the simulation shows that the upper Dirac cone surface state electrons are still found in the first quintuple atomic layer of Bi$_2$Se$_3$, but the corresponding lower Dirac cone electrons are absent, resembling a band gap as do our data (Fig. 1(c, right)).

\begin{figure}[t]
\includegraphics[width = 7.3cm]{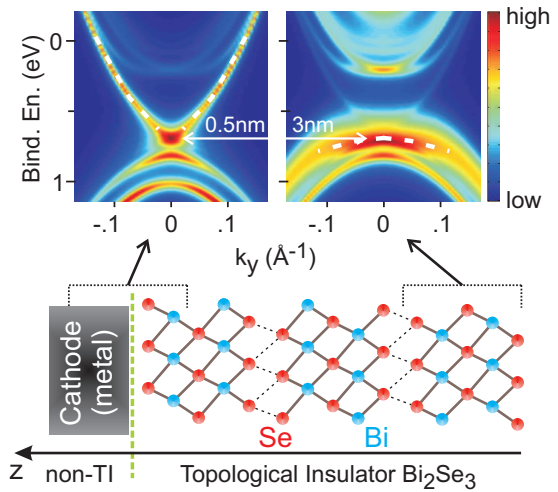}
\caption{\label{fig:SIMfig}{\bf{Combining electrostatics and electronic topology}}: A numerical simulation of copper-interfaced Bi$_2$Se$_3$ shows the partial density of low energy surface states at depths of 0-0.5nm and 2-3nm inside the crystal. (white dashed lines) Electrons from the upper Dirac cone of Bi$_2$Se$_3$ remain in the outermost 0.5nm, and electrons in the lower Dirac cone are located 1-2nm deeper in the crystal.}
\end{figure}

Looking two nanometers deeper inside the simulated crystal, we see that the missing in-gap electrons are part of a blurred continuum of relatively three-dimensional (3D) states derived from the bulk valence band. (Fig. \ref{fig:SIMfig}). Because these electrons are deeper in the crystal than the $\sim$1nm photoemission penetration depth, photoemission measurements will show a void or ``gap" near the D0 Dirac point. This phenomenon has been modeled previously in Ref. \cite{parMetal}, in which it is determined that a continuum of 3D states overlapping in energy and momentum with a TI surface state will cause the surface state dispersion to be reshaped with a gap-like appearance. Even though the electrons modeled in the right and left panels of Fig. \ref{fig:SIMfig} are physically displaced by several nanometers, there is sufficient wavefunction overlap between them to allow visible hybridization. No true 3D band gap is present \cite{parMetal}, however the experimental identification of such a gap-like surface feature in our data is significant because a similar overlapping of states is expected in many theoretically predicted topologically ordered metals such as half-Heusler compounds \cite{HeuslerDiscovery}. Reducing the band bending potential by 50$\%$ in our model causes the gap-like feature to disappear \cite{SMo}, consistent with recent measurements on limited-bias nanoscale N-N interfaces \cite{PanNN} and the absence of a gap after 0.25ML deposition ($\Delta_\mu$=0.33eV) in Fig. \ref{fig:PNfig}(a).

Taken collectively, the P-N interface electronic structures reported here reveal experimental ways in which the interaction of bulk and surface electrons reshapes surface band structure, and are consistent with numerical simulations in which the symmetry inversion of topological order generates a phenomenon that strongly resembles convential Rashba spin-splitting. Our experiment further suggests a characteristic lineshape of Dirac point interfaces, and how it can be understood in terms of the free path length (momentum broadening) of surface state electrons.

We thank  A. Balatsky, H. Dil, L. Fu, P. Ghaemi, and P. Hofmann for discussions. M.Z.H. acknowledges Visiting Scientist support from LBNL. M.Z.H. acknowledges additional support from the A. P. Sloan Foundation.

%\textbf{Methods summary:}
%Angle resolved photoemission spectroscopy (ARPES) measurements were performed at the Advanced Light Source beamlines 10 and 12 using 35-65 eV photons with better than 20 meV energy resolution and overall angular resolution better than 1$\%$ of the Brillouin zone (BZ). Samples were cleaved \emph{in situ} and typically measured at T$\lesssim$15$^o$K, in a vacuum maintained below 8$\times$10$^{-11}$ Torr. Samples deposited with In were studied at a higher temperature range of T=80$^o$ to 100$^o$K. Fe atoms were deposited using an e-beam heated evaporator at a rate of approximately 0.1$\AA$/minute, and Cu atoms were deposited at a rate of 0.4$\AA$/minute. A quartz micro-balance supplied by Leybold-Inficon with sub-Angstrom sensitivity was used to calibrate the deposition flow rate, and a factor of 2.5$\AA$/ML was used to estimate monolayer depth. Surface charge at low adatom coverage is estimated from the expected +1 ionization state of copper on the cleaved [111] Se surface \cite{WrayCuBiSe}. Adsorption of NO$_2$ molecules on Bi$_2$Se$_3$ was achieved by controlled \emph{in situ} exposures under static flow mode, with care to minimize photon exposure of the adsorbed surface. Crystal growth of large single crystals of Ca$_x$Bi$_{2-x}$Se$_3$ and Cu$_x$Bi$_2$Se$_3$ and numerical techniques used to simulate the perturbed TI surface are described in the SI.

\end{bibunit}

\

\begin{bibunit}
\begin{center}{\large \bf Supplemental Material for: Electron behavior in topological insulator based P-N overlayer interfaces}\end{center}
\renewcommand{\thefigure}{S\arabic{figure}}
\setcounter{figure}{0}
\renewcommand{\theequation}{S\arabic{equation}}

\subsection{SM I. Experimental Materials and Methods}

Spin-integrated angle-resolved photoemission spectroscopy (ARPES) measurements were performed with 35 to 65 eV photons on beamlines 10 and 12 at the Advanced Light Source, both endstations being equipped with a Scienta hemispherical electron analyzer (see VG Scienta manufacturer website $\langle$http://www.vgscienta.com/$\rangle$ for instrument specifications). The typical energy and momentum resolution were better than 20 meV and 1$\%$ of the surface BZ, respectively. Samples were cleaved \emph{in situ} and typically measured at T$\lesssim$15K, in a vacuum maintained below 8$\times$10$^{-11}$ Torr. Samples deposited with In were studied at a higher temperature range of T=80 to 100K. High quality single crystals of Ca$_x$Bi$_{2-x}$Se$_3$ were grown by a process of two-step melting, as described in Ref. \cite{HorCa}.

Copper, indium and iron atoms were deposited using an e-beam heated evaporator at a rate of approximately 0.4$\AA$/minute for Cu and In and 0.1$\AA$/minute for Fe. A quartz micro-balance supplied by Leybold-Inficon with sub-Angstrom sensitivity was used to calibrate the deposition flow rate prior to each experiment, leading to the possibility of some ($\sim$20$\%$) inconsistency in calibration between separate samples. Spectra of the adsorbed surfaces presented in the main paper were taken within minutes of opening the photon shutter to minimize potential photon induced charge transfer and desorption effects. No change was observed in the Cu-deposited surface energetics upon deliberately heating to 100K and exposing the sample to the photon beam for 30 minutes, suggesting that the surface was stable under experimental conditions.

Due to the very light coverage used, monolayer deposition depth is defined in terms of the number of adatoms per surface unit cell of Bi$_2$Se$_3$, with a large in-plane hexagonal lattice constant of approximately 4.15$\AA$. The area of a surface unit cell is (4.14$\AA)^2\times\sqrt{3}$/2=14.8$\AA^2$, and crystalline Fe and Cu have nearly identical densities of 85 atoms/nm$^3$, giving a naive conversion of 85 atoms/nm$^3$$\times$0.148nm$^2$= 12.6 ML/nm for each nanometer of calibrated deposition depth on the quartz balance. This value refers to the ratio of monolayer coverage on Bi$_2$Se$_3$ to the depth after an equal deposition time on the quartz nanometer in units of nanometers. Further, based on the change in surface energies as Cu is initially deposited, we estimate the sticking coefficient of copper on the chemically stable selenide surface to be S$_{Se}$$\sim$$\frac{1}{3}$, similar to the low temperature sticking coefficient for Cu on a comparably textured oxide surface \cite{CuEpitaxy}. Significantly larger sticking coefficients are incompatible with the observed shift in band energies near the crystal surface under light deposition, assuming that copper resides on the Se surface and has +1 valence. Normal-incidence copper growth on the chemically reactive surface of a quartz microbalance is generally assumed to have a sticking coefficient of unity, giving a final calibrated deposition rate of 1/(S$_{Se}$$\times$12.6 ML/nm)$\sim$2.4$\AA$/ML. This calibration is consistent with the decay of photoemission intensity from the outermost surface state bands as Cu is added. If we assume an ARPES penetration depth into the Cu layer of roughly 0.6nm at h$\nu$=65eV incident photon energy and 1nm at 35eV \cite{univCurve}, the decay of intensity 200meV above the original surface Dirac point is consistent with a thickness of 0.10$\pm$0.02 nm/ML for the first monolayer of deposition, which is reasonable for a single atomic layer of Cu.

For thicker layers such as the nominal ``2.2ML" measurements, coverage is likely greater, because the sticking coefficient will change as a Cu layer forms.  If we assume that the sticking coefficient changes linearly with coverage and reaches the sticking coefficient of the Cu-coated quartz microbalance after a number density of atoms equivalent to 2 Angstroms of bulk copper has been deposited, the actual coverage labeled as ``2.2ML" in the main text will be a number of Cu atoms equivalent to 2.35 ML of close packed [111] Cu planes (counting 0.2 nm/ML for bulk Cu).

%Cu has a sticking coefficient of 1 on Cu[111]:
%Ion solid surface interactions in ionized copper physical vapor deposition
%X.-Y. Liu *, a, M.S. Dawb,1, J.D. Kressc, D.E. Hansonc, V. Arunachalamb, D.G. Coronellb,2, C.-L. Liud,
%A.F. Voterc
%X.-Y. Liu et al. / Thin Solid Films 422 (2002) 141–149

Adsorption of NO$_2$ molecules on Ca$_x$Bi$_{2-x}$Se$_3$ was achieved by controlled exposures to NO$_2$ gas (Matheson, 99.5$\%$). The adsorption effects were studied under static flow mode by exposing the cleaved sample surface to the gas for a certain time then taking data after the chamber was pumped down to the base pressure. As with Cu, In and Fe deposition measurements, spectra of the NO$_2$ adsorbed surfaces were taken within minutes of opening the photon shutter to minimize photon exposure related effects.

\begin{figure*}[t]
\includegraphics[width = 10cm]{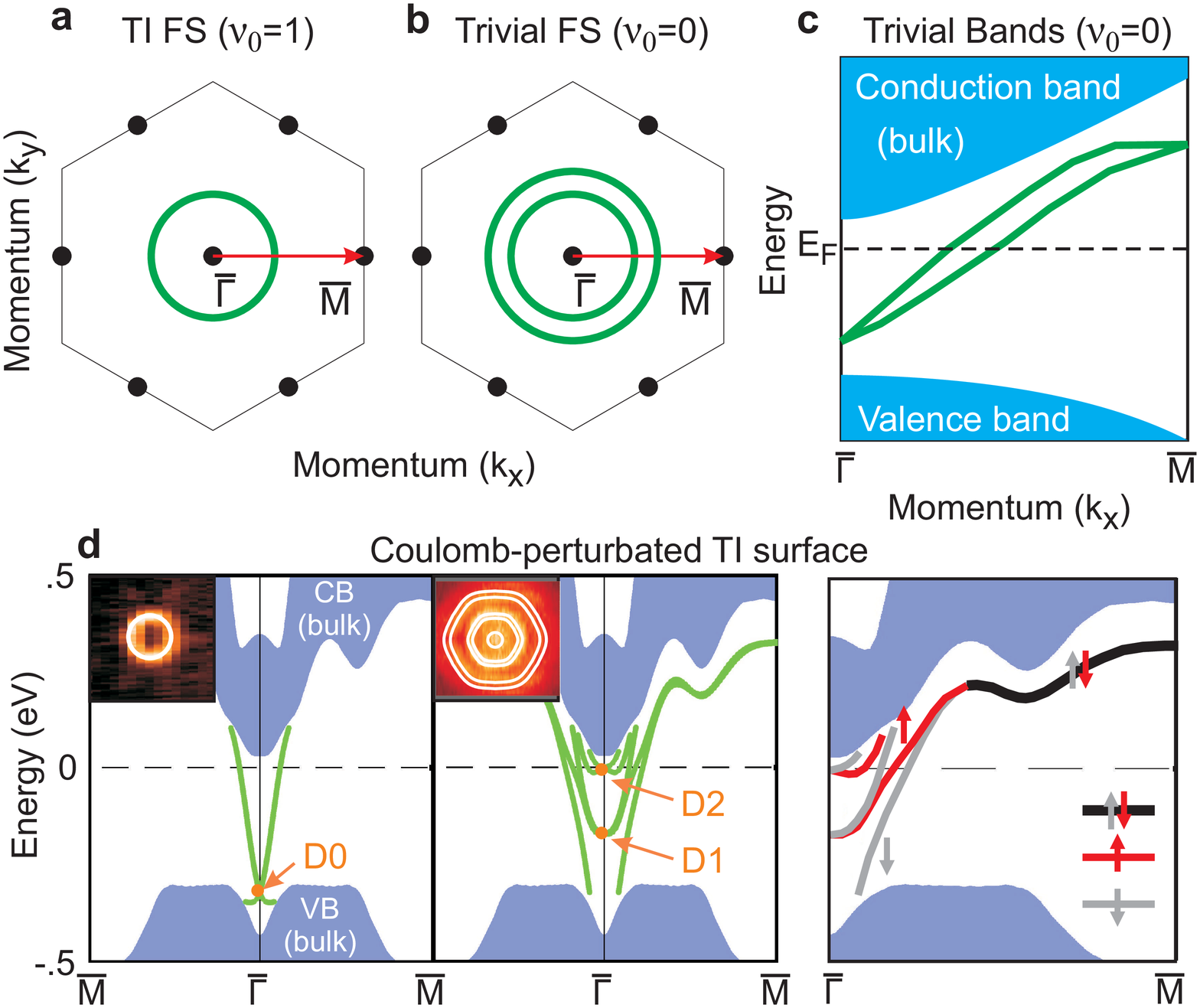}
\caption{\label{fig:TIconnect}{\bf{Topological insulator surface band connectivity}}: (a-b) Fermi surface diagrams are shown for a topological insulator and a non-topological insulator material. (c) A band structure diagram illustrates that at a topologically trivial surface, there are always an even number of surface state crossings between the high symmetry $\overline{\Gamma}$- and $\overline{M}$-points within the bulk band gap (here two). (d) Band structure diagrams show the GGA-predicted surface states for (left) vacuum-suspended Bi$_2$Se$_3$, and (middle, right) Bi$_2$Se$_3$ with positive charge on the surface. Details of the calculation can be found in Ref. \cite{WrayFe}. Experimental Fermi surfaces for cathode-deposited Bi$_2$Se$_3$ are shown as insets. (right) The GGA simulation shows that topological insulator surface states become doubly degenerate (thick black line) at large momentum.}
\end{figure*}

\subsection{SM II. Topological connectivity of surface state band structure}

Topological insulator surfaces can be identified by an odd number of surface states within the bulk band gap between certain high symmetry points in momentum space. In the case of Bi$_2$Se$_3$, an odd number of surface states must exist on any path through momentum space that connects the $\overline{\Gamma}$- and $\overline{M}$-points within the band gap. Diagrammatic examples of topological insulator and topologically trivial band structures are shown in Fig. \ref{fig:TIconnect}. Through surface adatom deposition (Cu, In and Fe), we have observed the number of surface states crossing the Fermi level between $\overline{\Gamma}$ and $\overline{M}$ changing from one to three to five, preserving this odd number rule (Fig. \ref{fig:TIconnect}(d)). Exceptions to the rule can be introduced if time reversal symmetry is broken by magnetic perturbations. Furthermore, surface states that are made degenerate with bulk electronic states (e.g. via band bending) are not explicitly topologically protected \cite{TIbasic}, as demonstrated by the parasitic metal band gap discussed in the main text.

One consequence of this topological definition is that, if the original upper Dirac cone (D0) is shifted downward in energy so that it no longer connects to the conduction band (defined by energetics deep in the crystal) as in Fig. \ref{fig:TIconnect}(d,right), the D0 upper Dirac cone must merge with another topological surface (interface) state as has been observed in our data (Fig. 1(c) of the main text). The extrinsically bound surface states introduced by surface perturbations also intersect the momentum space Kramers points in Dirac points, and fit the topological definition of topological surface states \cite{TIbasic}. In Fig. \ref{fig:TIconnect}(d) and Fig. 2(e) of the main text, an electrostatically perturbed 12-layer GGA slab calculation (details in Ref. \cite{WrayFe}) is used to show connectivity of extrinsically bound surface states with the original Bi$_2$Se$_3$ Dirac cone across the full Brillouin zone.

In counting the number of surface states, it is important to know if each state is singly degenerate (one state) or doubly degenerate, counting as two. If the slope of a surface state approaching the Brillouin zone center is non-zero forming a Dirac point, then the lattice symmetries of Bi$_2$Se$_3$ dictate that that state must be singly degenerate \cite{FuHexagonal}. Empirically, we find that singly degenerate states, including the original surface Dirac cone of pristinely vacuum-cleaved Bi$_2$Se$_3$, become doubly degenerate at large momenta (e.g. Fig. \ref{fig:TIconnect}(d) and Fig. 1(c) in the main text), causing the outermost surface state Fermi surface rings to become doubly degenerate after lengthy cathode deposition (including Cu, In, Fe). For Bi$_2$Se$_3$ the topological insulator state comes about as a result of band structure effects (termed ``topological symmetry inversion") that occur in the Brillouin zone center, and the double degeneracy of surface states far from the topological inversion point likely represents the typical behavior of a topologically trivial surface-perturbed material.

\subsection{SM III. Intrinsic band broadening at a Dirac singularity}

The broadening of electronic states along momentum and energy axes is a well known phenomenon that must be treated in an unusual way at Dirac singularities. Energy broadening occurs because energy and time are conjugate variables, and the finite lifetime of electronic states due to scattering causes their width along the energy axis to be non-zero. By the same token, momentum (e.g. `k$_x$') and spatial variables (e.g. `x') are conjugate to one another, and states are broadened along the momentum axes because scattering and localization restrict the translational symmetry of single-particle electronic states. For a band with small curvature, energy and momentum broadening cannot be readily distinguished and it is sufficient to broaden only in energy in comparing between theoretical band structure calculations and data. Dirac point singularities are a special case for which energy and momentum broadening cause distinctly recognizable effects. As an example, it is well known that the momentum-integrated electronic density of states goes to zero at a Dirac point in the absence of self energy broadening. Momentum broadening does not effect the distribution of electronic states along the energy axis, and thus allows the density of states at the Dirac point to remain zero. Energy-axis broadening (imaginary self energy) causes the momentum-integrated density of states at the Dirac point energy to become non-zero, and to approach the Dirac point energy in a parabolic rather than linear fashion. Large momentum broadening can create energy dispersion curve (EDC) lineshapes that strongly resemble a band gap, as presented in the main text.

As an example of how the effect of momentum broadening can be calculated, we have chosen momentum and energy broadening parameters of $\Gamma_k$=0.0275$\AA^{-1}$ and $\Gamma_E$=0.012eV such that the resultant broadening vector is approximately normal to the Dirac cone in the 2+1 dimensional momentum/energy space and corresponds with the intrinsic width prior to broadening from experimental resolution. The fitted lineshape is then obtained by broadening a Dirac cone with velocity v$_D$=2.3eV-$\AA$ through Lorentzian convolution in momentum and energy (convoluting each point [k$_{x0}$,k$_{y0}$,E$_0$] with $I\propto\frac{1}{(k_x-k_{x0})^2+(k_y-k_{y0})^2+(\Gamma_k/2)^2}\times\frac{1}{(E-E_0)^2+(\Gamma_E/2)^2}$). This `intrinsically' broadened electron distribution is then truncated by the 15K Fermi function and finally Gaussian energy broadened by the $\sim$20meV experimental resolution to produce the red curve in Fig. 3(d) of the main text. The characteristic lineshape that emerges from momentum broadening near the Dirac point is not particularly sensitive to the detailed values of $\Gamma_k$ and $\Gamma_E$, so long as $\Gamma_k$ is taken to be the dominant component determining linewidths.

\begin{figure*}[t]
\includegraphics[width = 12cm]{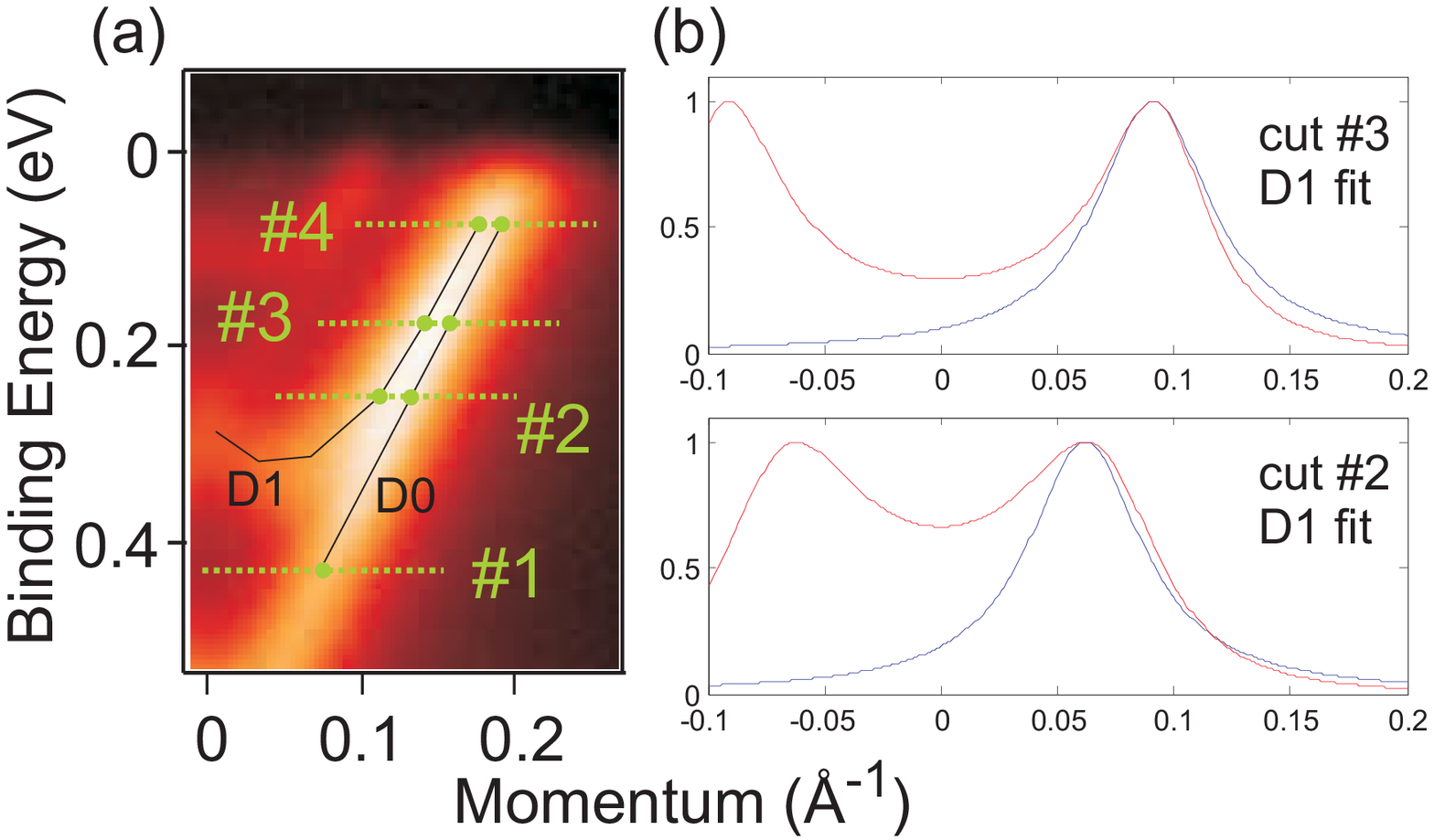}
\caption{\label{fig:BandFit}{\bf{Self energy broadening and the D1 lineshape}}: (a) Panel 2(b,left) from the main text is reproduced, showing energies at which the combined D0/D1 feature has been examined along the radial momentum axis. (b) The lineshape that would be expected for D1 from (gray) energy-axis broadening of the parabolic band is compared with (black) a fixed width Lorentzian ($\Gamma$=0.06$\AA^{-1}$). The parabolicity of D1 is approximated by fitting the black guide to the eye traced in panel (a).}
\end{figure*}

The choice of broadening function is also relevant for other parts of the photoemission spectrum, such as the band intersection fitted in Fig. 2(b) of the main text. Fig. 2(b) uses Lorentzian fitting functions, which does not account for effects from the detailed momentum dependence of the D1 band, such as the interplay of $energy-axis$ broadening with Rashba-like band parabolicity. A scenario in which energy broadening dominates for D1 would result in a very different fitting function for cut $\#$2 (see Fig. \ref{fig:BandFit}), and identifying the precise broadening function of D1 and D2 states may be an interesting subject for future studies.

\subsection{SM IV. Surface electron kinetics of $\rm Bi_2Se_3$ in presence of external potential using k.p theory}

The k.p Hamiltonian terms from Ref. \cite{FuNew,FuHexagonal} are adapted to include real space hopping parameters normal to the crystal surface, so as to allow simulation of electrostatic band bending. The reduced Hilbert space provided by an experimentally-based perturbative model has the advantage of allowing a crystal to be simulated up to arbitrary thickness, and of capturing the correct bulk kinetics, which are renormalized with respect to DFT-based calculations by a factor of $\sim$2 \cite{WrayPRB}. Unlike band bending models adopted from traditional materials such as GaAs \cite{HofmannRashba2DEG}, the modified k.p-derived model presented here accurately reproduces the surface state Dirac cone and topological effects associated with the topological symmetry inversion. However, because the model is defined perturbatively around the point of symmetry inversion in the 3D BZ center, it does not reproduce M-shaped dispersions in the valence band associated with k$_z$$\sim$$\pi$. Based on the symmetry analysis in Ref. \cite{FuNew,FuHexagonal}, the effective Hamiltonian for $\rm Bi_2Se_3$ near the Brillouin zone center for a quintuple layer is given by

\begin{eqnarray}
H_p = \left( \begin{array}{cccc}
k^2/m_1 & d+k^2/m_2 & ivk_x-vk_y & 0\\
d+k^2/m_2 & k^2/m_1 & 0 & -ivk_x+vk_y  \\
-ivk_x-vk_y & 0 & k^2/m_1 & d+k^2/m_2 \\
 0 & ivk_x+vk_y & d+k^2/m_2 & k^2/m_1 \end{array} \right),
\label{HH}
\end{eqnarray}

where $v$ is the Fermi velocity, $m_1$ and $m_2$ are orbital mass terms and the parameter $d$ induces a band gap by allowing spin-independent hopping between spatial states in the top and bottom halves of the quintuple layer. The physical context of this low-order perturbative model is that electrons near the Fermi level in Bi$_2$Se$_3$ primarily occupy p$_z$ orbitals of Se and Bi, with parity at the $\Gamma$-point defined by the relative phase (0 or $\pi$) for electrons on the top and bottom halves of a quintuple-layer. The perturbative Hamiltonian therefore explicitly considers spatial states on the top (columns 1 and 3) and bottom (columns 2 and 4) halves of a quintuple layer, that are differentiated by SU(2) spin (spin up in columns 1 and 2). The hopping between two adjacent quintuple layers is realized by

\begin{eqnarray}
  T = \left( \begin{array}{cccc}
t_{z1} & 0 & 0 & 0\\
t_z & t_{z1} & 0 & 0 \\
0 & 0 & t_{z1} & 0 \\
0 & 0 & t_z & t_{z1} \end{array} \right),
\label{T}
\end{eqnarray}
where $t_z$ is the hopping parameter and $t_{z1}$ breaks symmetry between z-axis dynamics of the conduction and valence bands.
To illustrate the single-Dirac-cone topological surface states, we choose the parameters to be : $m_1= 0.125 ~\rm eV^{-1}$-$\rm \AA^2$, $m_2= -0.04 ~\rm eV^{-1}$-$\rm \AA^2$, $d=-0.22$  eV, $v=2.5$ eV-$\rm \AA$, $t_z = 0.37$ eV, $t_{z1} = -0.045$ eV, matching the experimental band gap and conduction band mass.

\begin{figure}[htp]
\centering
\hskip-0.1 cm
\includegraphics[width=8cm]{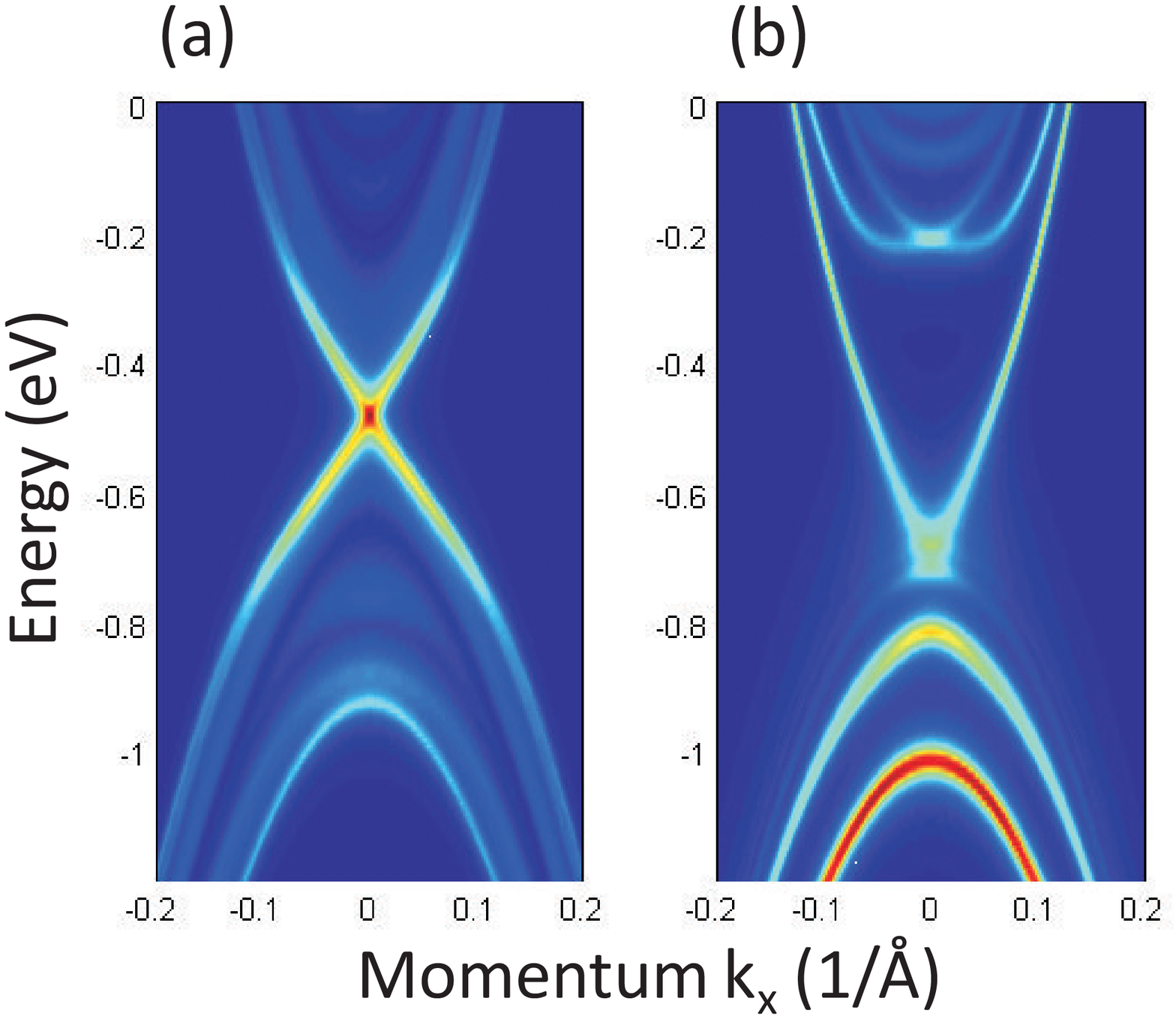}
\caption{ (a) Spectral weight from the top 1/2 quintuple-layer (top 0.5nm of the simulated crystal) without extra surface potentials. Binding energies are shifted to resemble the chemical potential of N-Type superconducting Cu$_{0.12}$Bi$_2$Se$_3$ (b) Weighted intensity from the first few layers with extra surface potentials to simulate Cu deposited ARPES intensity for P-type Bi$_{1.995}$Ca$_{0.005}$Se$_3$.}
\label{specwt}
\end{figure}

Fig. \ref{specwt}(a) displays the result of our calculation of spectral weight from the first 1/2 layer (0.5nm), solved for this Hamiltonian. The Dirac-cone like surface states are clearly seen inside the bulk gap.

Fig. \ref{specwt}(b) displays the result of a calculation of spectral weight for $\rm Bi_2Se_3$ with a Cu film deposited on the surface, described by surface potentials defined in the next section (Section SM V). It is also well known that the ARPES intensity decays with the distance from the surface. Keeping this in mind, we combine the contributions from first few layers to approximately match the expected experimental penetration depth of $\lesssim$1nm \cite{univCurve}. The intensity displayed in  Fig. \ref{specwt}(b) is a combination of 50\% contribution from the surface 1/2 quintuple-layer (top 0.5nm), 25\% contribution from the second layer, 12.5\% contribution from the third layer and 12.5\% contribution from the fourth layer.
%For further clarification we show the individual intensities from first two quintuple-layers in Fig. \ref{layers}.
%Here the surface is labeled as `layer 1' and three consecutive layers below the surface are labeled as `layer 2',`layer 3' and `layer 4', respectively.

\subsection{SM V. Estimating the surface potential after heavy cathode deposition}

Summing over the Luttinger count of all visible surface states shows that the total number of surface band electrons after a nominal deposition of 2 Cu atoms per surface unit cell on P-type Bi$_2$Se$_3$ is still only $\sim$0.1e-/BZ, giving a good estimate of the total number of electrons present near the surface with symmetries derived from the bulk conduction band. The nominal bulk carrier (hole) density from Hall measurements is approximately 10$^{-19}$e$^+$/cm$^3$=0.002e$^+$/BZ for the 3D Brillouin zone, meaning that negatively charged dopants will screen approximately 0.002 surface electrons per quintuple layer ($\sim$0.95nm) of depth. The depth of the depletion region can therefore be approximated by the ratio of the surface charge density ($\rho_{Cu^+}$+$\rho_{SS}$) and the bulk dopant density. ($\frac{\rho_{Cu^+}-0.1e^-/BZ}{0.002e^-/nm}$, assuming a large bulk band mass, which is accurate for the valence band).

Although Cu adatoms act as charge donors, they are not expected to approach a pure +1 ionization state after the chemical potential has shifted by several hundred meV. For the case of a single fully ionized monolayer of Cu (1 Cu$^+$ per surface BZ) the carrier depletion zone would have a depth of (1-0.1)/0.002=450nm, which is completely non-physical given the strong electric field that will be present at the surface (0.74 V/nm, even ignoring changes in the Bi$_2$Se$_3$ dielectric constant).

\begin{figure*}[htp]
\centering
\hskip-0.1 cm
\includegraphics[width=14.5cm]{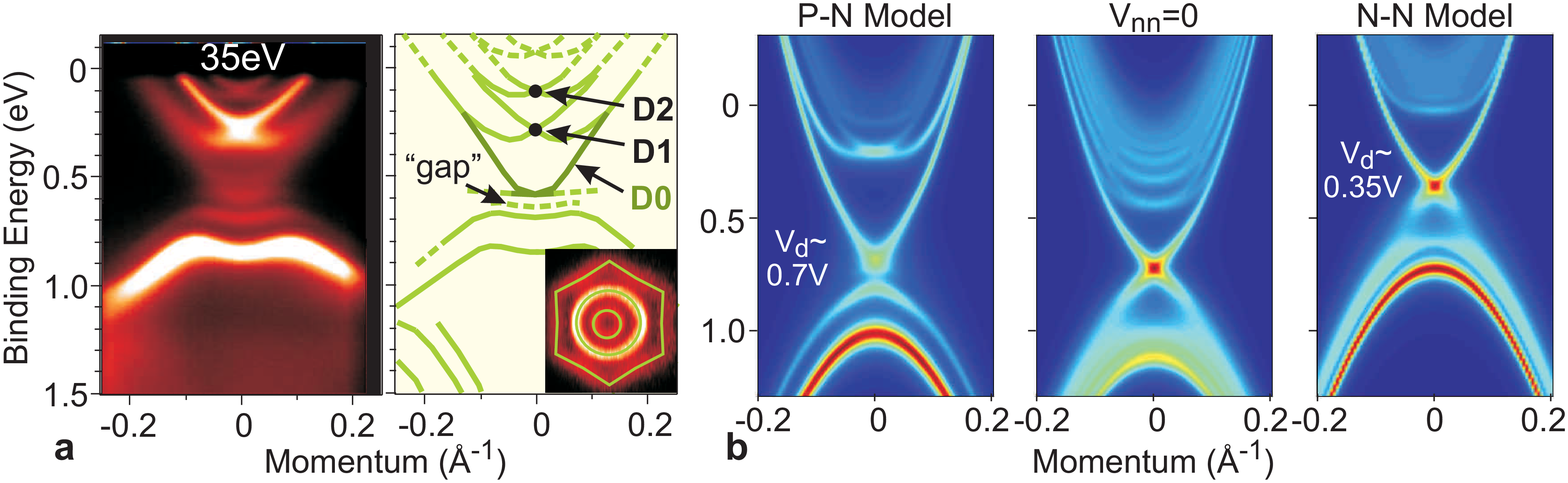}
\caption{\textbf{The importance of near neighbor surface interactions:} (a) ARPES measurements of interface states on the surface of P-type Bi$_2$Se$_3$ are shown from Fig. 1(c) of the main text. (b) States within the top quintuple layer ($\sim$1nm) of Cu-deposited Bi$_2$Se$_3$ are modeled (left) using a strong near-neighbor surface interaction as described in the text, (center) assuming screening with only long-range Coulomb interactions and the experimental dielectric constant of $\epsilon_r$=113, and (right) using the model described in the text with potentials reduced by 50$\%$ (interface voltage drop V$_d$$\sim$0.35V). Reducing the surface potential by 50$\%$ causes it to resemble the largest surface biases generally achieved across N-N interfaces, and accurately reproduces the recent experimental observation that N-N overlayer interfaces with biases up to V$_d$$\sim$0.35V manifest energy-axis broadening of the lower D0 Dirac cone band but no gap at the Dirac point \cite{PanNN}.}
\label{CompareVnn}
\end{figure*}

To reconcile the observed shift in surface band energies ($\sim$0.7eV) with the screening mechanisms at play, it is necessary that the total surface charge of Cu atoms be only $\lesssim$0.2e$^+$ per surface unit cell. However the kind of spectrum that appears in response to such a perturbation (middle panel in Fig. \ref{CompareVnn}(b)) is qualitatively incompatible with the measured ARPES spectrum in three respects. The points of incompatibility are: 1. The number of new bands derived from the valence and conduction bands is incorrect. 2. Spacing between bands at the surface is far too small, and no gap is expected to appear at the `D0' Dirac point. 3. Dirac velocities of new bands (Rashba splitting) are far too small. All of these effects indicate that a typical screening model (e.g. MTFA or Poisson-Schrodinger methods) with no short range correction severely underestimates the slope of the potential near the interface and how strongly new (extrinsic) surface states are confined at the TI surface when the density of surface adatoms is large. To more closely describe the surface kinetics, it is necessary to increase the slope of the surface potential in the top quintuple layers, to represent the short range effects such as plasticity in the dielectric constant and the near-neighbor interaction with Cu atoms. The modeled surface potential we have chosen, including polynomial long range screening and local perturbations, is given as

\begin{eqnarray}
V=V_0+V_1 z-V_2 z^2+V_{nn}(z),
\label{Vsurf}
\end{eqnarray}

where the depth `z' is represented by an integer index of the top 34 half-quintuple layers (16nm), with V$_0$=-0.35 eV, V$_1$=0.017 eV, and V$_2$=0.0002 eV. The value of V$_2$ is fixed by the bulk carrier density through the Possion equation, V$_0$ is determined by the change in surface energies, and V$_1$ is obtained by minimizing the electrostatic energy. The near-neighbor surface energy `V$_{nn}$' is an additional perturbation of -0.41, -0.28, -0.19, -0.15 eV respectively for the first four 1/2-quintuple layers (2nm). The principle origin of this sharply sloped surface potential in the top two nanometers derives from basic junction physics, and is a universal source of uncertainty in band bending models. The dielectric constant is not well defined near an interface, and changes by a factor of $\sim$20-40 from a strong cathode (e.g. Cu) to a semiconductor, meaning that the slope of the potential near the surface should be up to 40 times stronger than it would be with the dielectric constant of just the semiconductor. Our parameters attribute approximately 50$\%$ of the potential to this strongly sloped region.

A recent study that uses a more basic quantum model neglecting topological order has applied a global correction to the dipole constant (changing $\epsilon_r$ from 113 to 70) to partially account for this effect \cite{HofmannRashba2DEG}. However a global adjustment of that sort is relatively coarse and misses the key element that the potential gradient is only strongly enhanced in extremely close proximity to the interface. Furthermore, the origin of the large Rashba-like spin splitting observed in our model (and present in our data and first principles DFT-based simulations) is the symmetry inversion underlying topological order, and will be missed in any model that does not treat Bi$_2$Se$_3$ as a topological insulator.

This analysis of our data and recent literature suggests that the surface state dynamics after $\gtrsim$1ML of deposition (e.g. new bands and Rashba splitting) are almost entirely determined by near-neighbor interactions at the interface rather than long-range Coulomb interactions. Bulk-like screening with a roughly unchanged Bi$_2$Se$_3$ dielectric constant ($\epsilon_r$$\sim$100) is important at depths beyond 2-3nm into the crystal, and appears to contribute approximately half of the total binding energy of the surface states, even though the z-axis potential gradient of bulk-like screening is too small to greatly effect the shape of surface bands beyond a rigid energy shift. When the surface potential is steep enough, hybridization between these bulk-like states at d$\gtrsim$2nm inside the crystal and shallower surface bands can create a gap-like feature in the surface band structure, consistent with the theoretical proposal in Ref. \cite{parMetal}.

\end{bibunit}

\end{document}